# The Power of Small LLMs in Geometry Generation for Physical Simulations.

Ossama Shafiq**\***, Bahman Ghiassi**\*** and Alessio Alexiadis**\***,

**\*Correspondence**: Ossama Shafiq (oxs391@student.bham.ac.uk), Bahman Ghiassi (b.ghiassi@bham.ac.uk) and Alessio Alexiadis (a.alexiadis@bham.ac.uk)

## Abstract

Engineers widely rely on simulation platforms like COMSOL or ANSYS to model and optimise processes. However, setting up such simulations requires expertise in defining geometry, generating meshes, establishing boundary conditions, and configuring solvers. This research aims to simplify this process by enabling engineers to describe their setup in plain language, allowing a Large Language Model (LLM) to generate the necessary input files for their specific application. This highly novel approach allows establishing a direct link between natural language and complex engineering tasks.

Building on previous work that evaluated various LLMs for generating input files across simple and complex geometries, this study demonstrates that small LLMs— specifically, Phi-3 Mini and Qwen-2.5 1.5B—can be fine-tuned to generate precise engineering geometries in GMSH format. Through Low-Rank Adaptation (LoRA), we curated a dataset of 480 instruction–output pairs encompassing simple shapes (squares, rectangles, circles, and half circles) and more complex structures (I-beams, cylindrical pipes, and bent pipes). The fine-tuned models produced high-fidelity outputs, handling routine geometry generation with minimal intervention. While challenges remain with geometries involving combinations of multiple bodies, this study demonstrates that fine-tuned small models can outperform larger models like GPT-4o in specialised tasks, offering a precise and resource-efficient alternative for engineering applications.

## 1. Introduction

Engineers across industries rely heavily on simulation platforms like COMSOL (COMSOL AB, 2024) and ANSYS (ANSYS, 2024) to model and optimise complex processes. However, setting up these simulations presents a significant barrier – engineers must manually create precise input files defining geometries, meshes, boundary conditions, and solver configurations, or rely on graphical user interfaces (GUIs) which are not always available especially in open-source multi-physics simulations tools. This process is time-consuming, requires specialised expertise, and is prone to errors.

The automation of geometry generation, in particular, represents a crucial step toward making simulation tools more accessible to a broader engineering audience, allowing them to focus more on interpreting results rather than configuring simulations.  In educational settings, automation helps lower technical barriers that often discourage students from engaging with simulation tools. When students do not have to wrestle with the intricate details of input file creation and geometry setup, they can devote more cognitive resources to understanding the underlying engineering principles. This reduction in extraneous cognitive load is consistent with cognitive load theory, which posits that learning is optimised when unnecessary processing is minimised (Sweller et al., 2019).As a result, educators can design more effective instructional experiences that emphasise conceptual understanding and analytical skills over the mechanics of simulation configuration.

For computational research specifically, it is observed that automation of model generation accelerated iterative design processes, allowing researchers to explore larger parameter spaces and achieve more robust conclusions(Carta, 2020). Additionally, research highlights that language-model-driven automations in engineering education remove technical gatekeeping that historically disadvantaged students from non-traditional backgrounds, making advanced simulation techniques more accessible across diverse student populations (Filippi and Motyl, 2024).

Large Language Models (LLMs) have demonstrated remarkable capabilities in translating natural language descriptions into structured outputs, suggesting their potential to bridge this gap between human intent and technical implementation. Recent studies have shown promising results in applying LLMs to engineering tasks, from code generation to computer-aided design (CAD) (Filippi and Motyl, 2024b; Zhong et al., 2023). These applications leverage several key concepts, including prompt engineering and model-driven engineering (MDE). Prompt engineering enables few-shot learning, allowing models to generalise effectively from limited examples (Gu et al., 2022; Brown et al., 2020), while MDE principles help bridge the gap between high-level specifications and practical implementations (Gaydamaka, 2019; Butting et al., 2020).

However, applying general-purpose LLMs to specialised engineering tasks presents significant challenges, particularly when generating code for specialised tools like GMSH (Geuzaine and Remacle, 2009), an open-source finite element mesh generator that provides a scripting interface for creating complex engineering geometries (Rios et al., 2023). Our previous work revealed that without domain-specific training, even large LLMs struggle to consistently generate precise geometric shapes that adhere to engineering standards like GMSH syntax (Alexiadis and Ghiassi, 2024, Shafiq et al., 2025). These models often produce outputs that fail to meet the nuanced requirements of engineering design (Hu et al., 2021; Han et al., 2024). While some researchers have explored fine-tuning LLMs for CAD tasks (Sanghi et al., 2023; Wu et al., 2023), the focus has primarily been on larger models, which require substantial computational resources.

This study builds on our previous findings by investigating an alternative approach: the systematic fine-tuning of small LLMs—specifically Phi-3 Mini and Qwen-2.5 1.5B—for engineering geometry generation. Smaller models are critical for practical deployment in engineering contexts, as they require significantly fewer computations resources while potential offering faster inference times and easier integration into existing workflows. Our hypothesis is that these smaller models, when properly adapted using targeted datasets and training techniques, can match or exceed the performance of larger generalist models while being more accessible and efficient. Through careful dataset curation and specialised fine-tuning methods, we aim to enhance these models' ability to generate precise GMSH-compliant geometries from natural language descriptions. While our primary geometries involving a single body, we also examine the models' capabilities in handling geometries involving combinations of multiple bodies within a single prompt—a scenario that tests the limits of their contextual understanding.

In this paper, we first detail our approach to dataset creation and model fine-tuning, describing the techniques that enable small LLMs to generate precise engineering geometries. We then present comprehensive results that demonstrate these models' capabilities across a range of geometry types, from simple shapes to complex structures. Our analysis explores both the successes and limitations of small LLMs in this domain, leading to insights about their practical implementation in engineering workflows. We conclude by examining the broader implications of our findings for the future of automated engineering design and simulation setup.

## 2. Previous Work

### 2.1 Efficient LLM Adaptation Techniques

Recent research has demonstrated that language models can be effectively fine-tuned for specialised engineering tasks. Parameter-efficient fine-tuning methods such as Low-Rank Adaptation (LoRA) (Hu et al., 2021) have emerged as practical approaches for adapting pre-trained models to domain-specific applications without the computational burden of full fine-tuning. These techniques modify only a small subset of model parameters while preserving most of the pre-trained weights, effectively mitigating the problem of catastrophic forgetting where models lose their general capabilities when adapted to new tasks.

For resource-constrained applications, Dettmers et al. (2023) introduced QLoRA, which combines 4-bit quantisation with LoRA to dramatically reduce memory requirements during fine-tuning. This approach is particularly relevant for our work as it enables adaptation of small models like Qwen-2.5 14B to specialised tasks such as engineering geometry generation, making deployment feasible in typical engineering computing environments.

### 2.2 LLMs in Engineering Applications

The application of LLMs to engineering workflows represents an emerging field with significant potential. Alexiadis and Ghiassi (2024) demonstrated a method for integrating LLMs with geometry/mesh generation software and multiphysics solvers, allowing users to describe simulations in natural language. Their work established the feasibility of using LLMs for engineering simulation setup, but primarily relied on larger models through API access, which presents limitations for widespread adoption in engineering practice.

Building on this, Shafiq et al. (2025) explored the capabilities of various LLMs specifically for generating GMSH geometry code from natural language descriptions. Their research identified significant gaps in the performance of even state-of-the art models when handling complex engineering geometries, highlighting the need for domain-specific adaptation. Similar efforts include FLUID-GPT (Yang et al., 2023) for particle trajectory prediction and MyCrunchGPT (Kumar et al., 2023) for scientific machine learning applications, further demonstrating the growing interest in applying LLMs to specialised engineering tasks.

### 2.3 CAD and Geometry Generation

The automation of CAD (Computer Aided Design) and geometry generation for engineering simulations presents unique challenges distinct from general CAD automation. While there has been progress in natural language interfaces for CAD systems, such as Alrashedy et al.'s (2024) work with

AutoCAD, these approaches typically focus on the broader design process rather than the specific requirements of simulation-ready geometry formats like GMSH.

Wu et al. (2021) demonstrated with DeepCAD that treating CAD operations as sequential commands can enable automated geometry generation, though their approach used specialised neural networks rather than adapting general-purpose LLMs. This distinction is crucial, as engineering simulation workflows often require both natural language understanding and precise adherence to domain-specific syntax, a combination that general CAD automation approaches have not fully addressed.

### 2.4 Current Challenges and Research Gaps

While general-purpose LLMs show promise in understanding engineering concepts, they often fail to produce the precise, syntactically correct geometry definitions required by simulation tools like GMSH.

Current approaches typically rely on large models that require significant computational resources, limiting their practical deployment in routine engineering workflows. There is a clear need for more efficient solutions that maintain high performance while reducing resource requirements.

Most existing solutions address either geometry generation or simulation setup in isolation, creating a disconnect that requires manual intervention. An integrated approach that can reliably generate simulation-ready geometry code from natural language would significantly streamline engineering workflows.

Our work directly addresses these gaps by fine-tuning small, efficient LLMs specifically for engineering geometry generation in GMSH format. By focusing on this specialised task and employing parameter-efficient techniques, we demonstrate that properly adapted smaller models can provide the precision required for engineering applications while being more accessible for routine use.

## 3. Methodology

Our research methodology followed a systematic five-phase approach to evaluate and enhance LLM performance in engineering geometry generation.

1. Conducted **baseline testing of selected models** to establish performance benchmarks. This involved evaluating three small LLMs (Phi-3 Mini (Abdin et al., 2024), Qwen-2.5 1.5B, and Qwen-2.5 14B (Qwen Team, 2025)) and comparing their capabilities with GPT-4o (Open AI, 2024), a larger model, in generating basic GMSH-compliant geometries.
2. Developing a **specialised dataset** for training by creating automated scripts to generate instruction-output pairs, implementing validation protocols to ensure geometric accuracy and GMSH compliance.
3. Implementing **fine-tuning procedures** using LoRA adaptation techniques to Phi-3 Mini and Qwen-2.5 1.5B, and QLoRA for the larger Qwen-2.5 14B model to manage computational resources.

4. Conducted comprehensive **performance validation** across multiple dimensions, testing models on single-geometry generation tasks and evaluating both GMSH syntax compliance and geometric accuracy.
5. Extended our investigation to more challenging scenarios through experiments involving **geometries with multiple bodies**, comparing the capabilities of both fine-tuned and non-tuned models.

3.1 Model Selection and Architecture

Our selection deliberately spans multiple architectural scales to investigate whether targeted fine-tuning of smaller models can match or exceed larger architectures in specialised engineering domains *(Table 1)*.

*Table 1.* Details on the architecture of the models selected.

| Model | Parameters (billion) | Architecture | Context Window | Key Features |
|---|---|---|---|---|
| **Phi-3-Mini** (Abdin et al., 2024) | 3.8B | Decoder-only Transformer | 4K tokens | Trained on 3.3T tokens including synthetic "textbook-quality" material |
| **Qwen-2.5 1.5B** (Qwen Team, 2025) | 1.5B | Standard Transformer | 128K tokens | Optimised for inference speed and memory efficiency |
| **Qwen-2.5 14B** (Qwen Team, 2025) | 14.7B | Standard Transformer | 128K tokens | ~40 query attention heads with 8 key/value pairs |
| **GPT-4o** (Open AI, 2025) | Not disclosed | Not disclosed | 128K tokens | State-of-the-art model used as performance benchmark |

3.2 Database Development

The dataset was designed to provide comprehensive coverage of geometry types while maintaining equal representation across parameter variations to avoid biasing the model towards specific configurations. To ensure systematic coverage of the parameter space, the following methodologies were implemented:

- Template variety and structure *(Figure 1)*: We developed a set of 8 distinct instruction templates for each geometry type, varying in their structure and complexity. For example, templates for circle generation ranged from simple commands like "Create a gmsh script for a circle with radius X" to more detailed specifications like "Generate a gmsh script for a X unit circle with mesh size Y."

```python
class TemplateManager:
    def __init__(self):
        self.circle_templates = [
            "Create a gmsh script for a circle with radius {radius} {unit}",
            "Make a gmsh script for a {radius} {unit} circle",
            "Generate gmsh script: {radius} circle with mesh {mesh_size}"
        ]
        self.semi_circle_templates = [
            "Create a gmsh script for a semicircle with radius {radius} {unit} facing {orientation}",
            "Make a gmsh script for a {radius} semicircle facing {orientation}"
        ]
```

*Figure 1.* Example instruction templates for circle geometry, showing variety in structure and complexity.

- Parameter diversity controls *(Figure 2)*: Rather than random parameter selection, we implemented controlled variation of key geometric parameters. We evenly distributed dimensions between integer and floating-point values, balanced unit specifications between millimetres and centimetres, and equally divided examples between explicit coordinate positioning and default positioning at the origin.

```python
def generate_params(self, shape_type: str, template: str) -> Dict:
    # Alternates between integer and float values
    use_integers = random.choice([True, False])
    # Varies units between millimeters and centimeters
    unit = random.choice(["mm", "cm"])
    # Scales values based on unit choice
    base_value = 100 if unit == "mm" else 10
```

*Figure 2.* Parameter diversity control implementation showing distribution of dimensional parameters.

Quality control features *(Figure 3)*: Each instruction-output pair underwent a rigorous validation process consisting of: (a) GMSH syntax validation to ensure the generated code was executable, (b) geometric validation to verify that the output code produced the intended shape, and (c) parameter consistency checking to confirm that all requested parameters were correctly represented in the output code.

```python
def generate_template_variations(self, template: str, shape_type: str):
    # Checks for required parameters in templates
    requires_unit = "{unit}" in template
    requires_position = "{center}" in template
```

*Figure 3.* Quality control validation function for checking GMSH syntax and geometric correctness.

We created a dataset of 480 instruction output pairs (120 per geometry type) covering four categories of geometries selected to represent increasing complexity in engineering applications.
- Basic shapes (squares/rectangles): Fundamental elements in structural and mechanical engineering designs.
- Curved elements (circles/semi-circles): Essential for modelling components with curved surfaces.
- Structural components (I-beams): Representative of complex engineering structures with specific dimensional requirements.
- Advanced geometries (pipes/bent pipes): Complex 3D structures that require precise parameterisation.

For each geometry type, we developed Python scripts to automatically generate diverse instruction-output pairs The pairs were formatted in JSON structure to facilitate easy integration with the fine-tuning pipeline *(Figure 4)*. These scripts systematically produced varied geometric configurations according to our parameterisation rules, enabling efficient dataset creation with comprehensive parameter space coverage. We created multiple variations of instructions using different phrasings but requesting the same geometry, for example the same square geometry might be requested using different instruction formats:

- Instruction: A natural language description of the desired geometry (e.g., "Create a gmsh script for a square with side length 10 units centred at the origin")
- Output: The corresponding GMSH geometry (.geo) file content that generates the specified shape.

```
{
    "instruction": "Create gmsh geometry for I-beam: d=560, bf=518, tw=28, tf=78 mm",
    "output": "// I-beam parameters\nh = 560;  // Total height mm\nb = 518;   // Flange width mm\ntw = 28;   // Web thickness mm\ntf = 78; // Flange thickness mm\n\n// Define mesh size\nmesh_size = 1.0;\n\n// Bottom flange points\nPoint(1) = {0 - b/2, 0 - h/2, 0, mesh_size};\nPoint(2) = {0 + b/2, 0 - h/2, 0, mesh_size};\nPoint(3) = {0 + b/2, 0 - h/2 + tf, 0, mesh_size};\nPoint(4) = {0 + tw/2, 0 - h/2 + tf, 0, mesh_size};\nPoint(5) = {0 - tw/2, 0 - h/2 + tf, 0, mesh_size};\nPoint(6) = {0 - b/2, 0 - h/2 + tf, 0, mesh_size};\n\n// Web points\nPoint(7) = {0 - tw/2, 0 + h/2 - tf, 0, mesh_size};\nPoint(8) = {0 + tw/2, 0 + h/2 - tf, 0, mesh_size};\n\n// Top flange points\nPoint(9) = {0 + b/2, 0 + h/2 - tf, 0, mesh_size};\nPoint(10) = {0 + b/2, 0 + h/2, 0, mesh_size};\nPoint(11) = {0 - b/2, 0 + h/2, 0, mesh_size};\nPoint(12) = {0 - b/2, 0 + h/2 - tf, 0, mesh_size};\n\n// Define lines\nLine(1) = {1, 2}; Line(2) = {2, 3}; Line(3) = {3, 4};\nLine(4) = {4, 5}; Line(5) = {5, 6}; Line(6) = {6, 1};\nLine(7) = {5, 7}; Line(8) = {7, 8}; Line(9) = {8, 4};\nLine(10) = {8, 9}; Line(11) = {9, 10}; Line(12) = {10, 11};\nLine(13) = {11, 12}; Line(14) = {12, 7};\n\n// Define surfaces\nLine Loop(1) = {1, 2, 3, 4, 5, 6};        // Bottom flange\nLine Loop(2) = {4, 7, 8, 9};              // Web\nLine Loop(3) = {8, 10, 11, 12, 13, 14};   // Top flange\n\nPlane Surface(1) = {1};  // Bottom flange\nPlane Surface(2) = {2};  // Web\nPlane Surface(3) = {3};  // Top flange"
},

{
    "instruction": "Generate gmsh script: 2 circle with mesh 0.71",
    "output": "// Define parameters\n    radius = 2;\n    mesh_size = 0.71;\n    center_x = 0;\n    center_y = 0;\n\n    // Define points\n    Point(1) = {center_x + radius, center_y, 0, mesh_size};\n    Point(2) = {center_x, center_y + radius, 0, mesh_size};\n    Point(3) = {center_x - radius, center_y, 0, mesh_size};\n    Point(4) = {center_x, center_y - radius, 0, mesh_size};\n    Point(5) = {center_x, center_y, 0, mesh_size}; // Center point\n\n    // Define circle arcs\n    Circle(1) = {1, 5, 2};\n    Circle(2) = {2, 5, 3};\n    Circle(3) = {3, 5, 4};\n    Circle(4) = {4, 5, 1};\n\n    // Define surface\n    Line Loop(1) = {1, 2, 3, 4};\n    Plane Surface(1) = {1};"
}
```

***Figure 4.*** The instruction-output pairs in JSON format for I-beam and circle, as an example

The automation scripts incorporated parametric variations to ensure diversity in the dataset ***(Table 2)***.

*Table 2:* Specific parameter ranges used in dataset generation.

| Parameter Type | Range/Options | Details |
|---|---|---|
| Dimensional Variation | Integers: 1-100 units<br>Floats: 0.1-99.9 units | Square/rectangle sides: 1-100 units<br>Circle/semi-circle radius: 0.5-50 units<br>I-beam web height: 10-100 units<br>I beam flange width: 5-50 units<br>Pipe diameter: 1-50 units<br>Pipe length: 5-200 units |
| Positional Variation | Centre coordinates (x/y/z): -50 to +50 | Origin-centered: 0,0,0<br>Quadrant 1: 1-50, 1-50, 0-50<br>Quadrant 2: -50 to -1, 1-50, 0-50<br>Quadrant 3: -50 to -1, -50 to -1, 0-50<br>Quadrant 4: 1-50, -50 to -1, 0-50 |
| Orientation Variation | Angles: 0-360° in 15° increments | Rotation around (x/y/z-axis): 0°, 15°, 30°, …, 360° |
| Language Variation | 5-10 instruction templates per geometry type | Command style: "Create a…"<br>Request style: "I need a…"<br>Question style: "Can you generate a…"<br>Technical style: "Write GMSH code for a…"<br>Detailed style: "Generate a square with dimension…" |

### 3.3 Fine-Tuning Procedure

We fine-tuned three small language models—Phi-3 Mini, Qwen-2.5 1.5B, and Qwen-2.5 14B—specifically for the task of generating GMSH-compliant geometry code from natural language descriptions *(Appendix 1)*. The fine-tuning process utilised Low-Rank Adaptation (LoRA), a parameter-efficient method that allows for model adaptation while minimising computational requirements.

**Hyperparameter Optimisation**

To determine optimal fine-tuning configurations, we conducted a systematic hyperparameter search based on ranges commonly used in prior LoRA fine-tuning studies (Hu et al., 2021; Menick et al., 2022; Han et al., 2024)- *Table 3* summarises the hyperparameters tested and their rationale:

*Table 3.* Hyperparameter search space for fine-tuning experiments.

| Hyperparameter | Values Tested | Rationale |
|---|---|---|
| LoRA rank | 4, 8, 16 | These values represent a common range in the literature. Hu et al. (2021) demonstrated that ranks 4-16 provide sufficient adaptation capacity for most tasks while remaining computationally efficient. Lower ranks (r=4) offer faster training but potentially limited expressivity, while higher ranks (r=16) provide greater adaptation capacity at increased computational cost. |
| LoRA alpha | 8, 16, 32 | Alpha controls the scaling of LoRA updates, thereby modulating how aggressively the model adapts during fine-tuning. Hu et al. (2021) demonstrate that starting with a simple heuristic—often setting alpha equal to the rank—is effective. In practice, however, many researchers adopt a slightly more aggressive approach by setting alpha to 2× the rank (e.g., using 8 when r=4 or 16 when r=8), which tends to balance rapid adaptation with computational efficiency. |
| Learning rate | 1.0e-05, 6.0e-05, 1.0e-04 | These AdamW optimizer learning rates span the range typically used for LoRA fine-tuning. Han et al. (2024) found rates in the 1e-5 to 1e-4 range effective for domain adaptation of language models. All implementations used linear warmup followed by cosine decay scheduling. |
| Training epochs | 1, 3, 5 | Unlike image models which often require many epochs, language model fine-tuning typically requires fewer epochs to avoid overfitting, especially on smaller datasets. Menick et al. (2022) demonstrated that 1-5 epochs are often sufficient for LLM adaptation tasks, with overfitting becoming a concern beyond that range. |

For each hyperparameter configuration, we trained the models and evaluated their performance on a validation set containing examples of all geometry types. The evaluation metrics included successful code generation rate, syntax correctness, and geometric accuracy. The optimal configurations determined from this process were implemented for the final model training, with results presented in *Section 4.1*.

For the smaller models, Phi-3 Mini and Qwen-2.5 1.5B, we implemented standard LoRA adaptation due to their manageable parameter counts. However, the larger Qwen-2.5 14B model required a more memory-efficient approach, for this model, we implemented 4-bit quantisation through

QLoRA. While this introduces a methodological difference, QLoRA maintains performance nearly identical to standard LoRA while significantly reducing memory usage, with differences in task performance typically less than 1% (Dettmers et al., 2023).

**Training Data Preparation**

The standardised prompt template format was applied at the fine-tuning stage to prepare the data into the required format ensuring consistent model input. *Figure 5* demonstrates the template structure for Phi-3-Mini, which uses special tokens to delimit instruction and code sections. For the Qwen-2.5 models (both 1.5B and 14B variants), we implemented a different template structure as shown in *Figure 6*, which uses the model's native instruction format and tokens to properly encapsulate user requests and assistant responses.

```
<|system|>
You are a helpful assistant.<|end|>
<|user|>
{prompt}<|end|>
<|assistant|>
```

*Figure 5.* Template format for Phi-3-Mini fine-tuning with special tokens delimiting input and output sections (Lightning AI, 2023).

```
<|im_start|>system
[System Message]<|im_end|>
<|im_start|>user
[User's Input]<|im_end|>
<|im_start|>assistant
[Assistant's Response]
```

*Figure 6.* Template format implementation for Qwen-2.5 models showing the use of model-specific instruction tokens that differ from those used with Phi-3-Mini (Lightning AI, 2023).

**Implementation Environment**

The fine-tuning process was implemented using LitGPT (Lightning AI, 2023), a lightweight and efficient framework designed for training and fine-tuning large language models. The LitGPT library provided a streamlined interface for integrating LoRA and QLoRA, enabling rapid experimentation and optimisation of hyperparameters. The process was executed on the Birmingham Environment for Academic Research (BEAR) high-performance computing cluster, utilising an NVIDIA A100 40GB GPU.

**Testing and Evaluation**

After the fine-tuning process, a series of seven distinct prompts were applied to generate GMSH scripts for a variety of geometries to test and evaluate the large language models *(Figure 7)*.

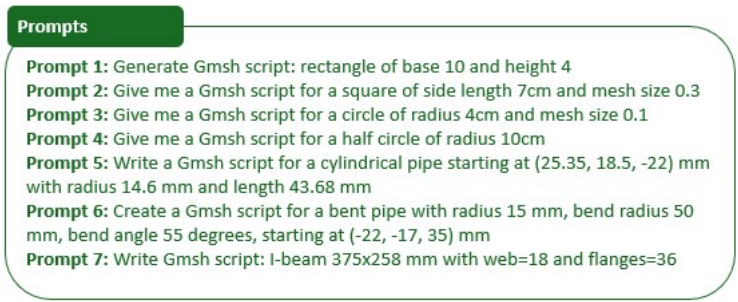

*Figure 7*. Series of prompts for testing and evaluating LLMs.

Each output was validated for GMSH compatibility based on the following criteria to produce geometry outputs as expected in **Figure 8**.

In addition to evaluating the performance on individual geometric shapes, we designed a series of tests specifically to assess the models' capabilities in handling geometries involving multiple bodies within a single prompt. These multi-body geometry tests were created to evaluate how well the models could manage spatial relationships and geometric interactions between different shapes. The multi-body geometry prompts **(Figure 9)** included scenarios such as 'a circle inside a square' and 'a circle next to a square,' requiring the models to correctly interpret relative positioning and generate appropriate GMSH code that would produce the specified composite structures.

*Table 4.* Evaluation criteria and scoring system for generated geometry code (Shafiq et al., 2025).

| Category | Subcategory | Max Points | Scoring Criteria |
|---|---|---|---|
| Geometry Fidelity | Shape Accuracy | 15 | Perfect match: 15 points<br>Minor deviations: 10 points<br>Noticeable alterations: 5 points<br>Major deviations: 0 points |
|  | Dimensional Accuracy | 15 | Within 1% tolerance: 15 points<br>Within 5% tolerance: 10 points<br>Within 10% tolerance: 5 points<br>10% error: 0 points |
| Simulation Parameters | Parameter Matching | 15 | Perfect match: 15 points<br>Minor discrepancies: 10 points<br>Significant differences: 5 points<br>Incorrect parameters: 0 points |
|  | Unit Consistency | 15 | Perfectly consistent: 15 points<br>Minor inconsistencies: 10 points<br>Major inconsistencies: 5 points<br>Completely inconsistent: 0 points |
| **Total Possible Score** |  | **60** |  |

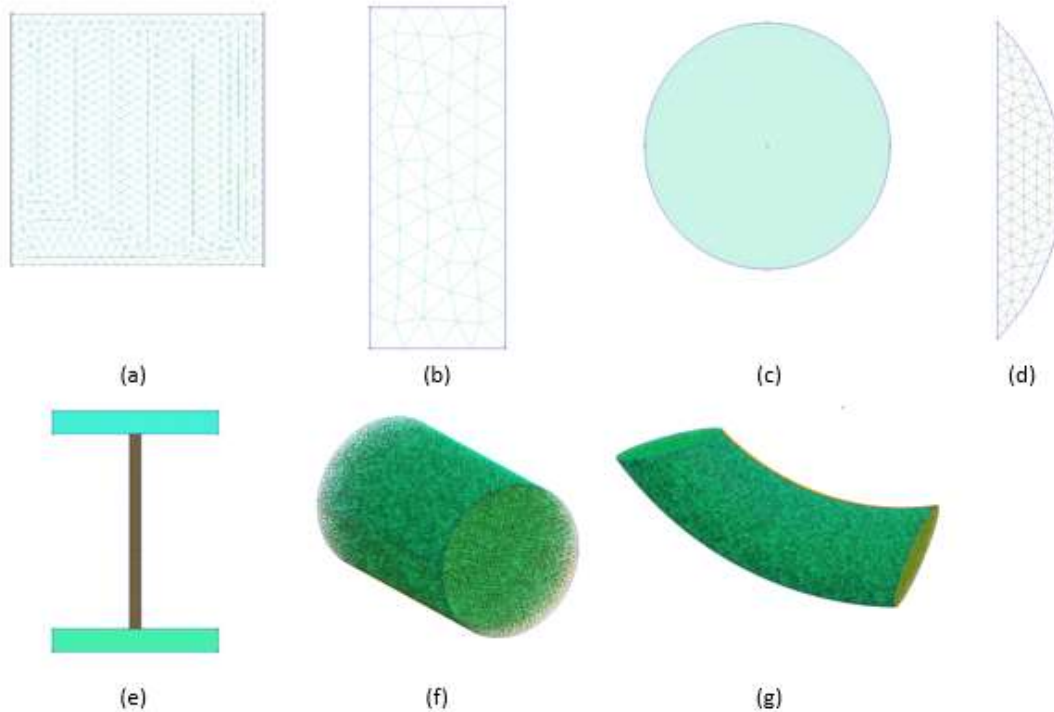

*Figure 8.* The expected visual representation of the square (a), rectangle (b), circle (c), semi-circle (d), I-beam (e), pipe (f) and bent pipe (g) geometries.

In additional experiments designed to evaluate the models' performance on more complex tasks, we prompted the models with a single input requesting multiple, distinct geometries simultaneously *(Figure 9)*.

**Prompt (a)**
Generate a GMSH script for a circle of radius 3cm inside a square of side length 7cm.

**Prompt (b)**
Give me a Gmsh script for a circle next to a square.

*Figure 9*. Series of prompts for evaluating the multi-geometry capabilities of LLMs.

## 4. Results & Analysis

### 4.1 Baseline Performance before Fine-Tuning

The initial evaluation of LLMs revealed significant variations in performance across different model sizes and architectures. The baseline assessment focused on four key criteria, as reflected in *Section 3.3,* including shape accuracy, dimensional accuracy, parameter matching, and consistent units, with scores ranging from 0 to 15 for each criterion *(Table 5).*

**Table 5.** Comprehensive table showing all scores across the seven geometries for four LLMs before fine tuning.

| LLM | Geometry Criteria | Square | Rectangle | Circle | Semi-Circle | Pipe | Bent Pipe | I-Beam |
|---|---|---|---|---|---|---|---|---|
| Phi-3-Mini | Shape Accuracy | 0 | 10 | 0 | 0 | 0 | 0 | 0 |
| | Dimensional Accuracy | 0 | 15 | 5 | 0 | 5 | 5 | 5 |
| | Parameter Matching | 0 | 5 | 0 | 0 | 0 | 0 | 0 |
| | Consistent Units | 5 | 10 | 0 | 0 | 0 | 0 | 0 |
| | TOTAL /60 | 5 | 40 | 5 | 0 | 5 | 5 | 5 |
| Qwen-2.5 1.5B | Shape Accuracy | 0 | 5 | 0 | 0 | 0 | 0 | 0 |
| | Dimensional Accuracy | 5 | 0 | 5 | 0 | 5 | 5 | 5 |
| | Parameter Matching | 0 | 5 | 0 | 0 | 0 | 0 | 0 |
| | Consistent Units | 5 | 5 | 5 | 0 | 5 | 0 | 0 |
| | TOTAL /60 | 10 | 15 | 10 | 0 | 10 | 5 | 5 |
| Qwen-2.5 14B | Shape Accuracy | 15 | 15 | 0 | 5 | 5 | 5 | 5 |
| | Dimensional Accuracy | 15 | 15 | 5 | 10 | 10 | 10 | 10 |
| | Parameter Matching | 10 | 10 | 0 | 5 | 5 | 0 | 5 |
| | Consistent Units | 15 | 15 | 5 | 5 | 10 | 5 | 10 |
| | TOTAL /60 | 55 | 55 | 10 | 25 | 30 | 20 | 30 |
| GPT-4o | Shape Accuracy | 15 | 15 | 15 | 15 | 10 | 5 | 10 |
| | Dimensional Accuracy | 15 | 15 | 15 | 15 | 15 | 10 | 15 |
| | Parameter Matching | 15 | 15 | 15 | 10 | 5 | 0 | 5 |
| | Consistent Units | 15 | 15 | 15 | 15 | 15 | 10 | 15 |
| | TOTAL /60 | 60 | 60 | 60 | 55 | 45 | 25 | 45 |

**Small LLMs (Phi-3 Mini and Qwen-2.5 1.5B)**

Both Phi-3 Mini and Qwen-2.5 1.5B demonstrated limited capabilities in their baseline state, with scores significantly lower than GPT-4o, which achieved perfect scores (60/60) for basic geometries as shown in *Table 6*. While GPT-4o exhibited strong performance even without fine-tuning, small models required significant improvement to reach comparable accuracy levels. Phi-3 Mini achieved its highest score with rectangles (40/60), but performed poorly across other geometries, scoring only 5/60 for most shapes. Qwen-2.5 1.5B showed slightly more consistent but still limited performance, with scores ranging from 5 to 15/60 across different geometries. The models frequently produced formatting and syntactical errors in GMSH scripts, while struggling with accurate geometric parameter definitions. Their handling of multiple geometry requests was notably inconsistent, indicating fundamental limitations in their baseline capabilities.

**Medium-Scale Model (Qwen-2.5 14B)**

Qwen-2.5 14B demonstrated markedly improved baseline performance compared to its smaller counterparts, particularly with basic geometries, though still below GPT-4o's capabilities for most geometries. While GPT-4o achieved perfect scores across basic shapes, Qwen-2.5 14B's strongest performance (55/60) for squares and rectangles still showed room for improvement, also showing moderate capability with pipes and I-beams (30/60) *(Figure 10)*. It maintained consistent dimensional accuracy across most shapes, though it still struggled with complex geometries, scoring lower on circles (10/60). This performance suggested that increased model size contributed to better baseline handling of geometric tasks.

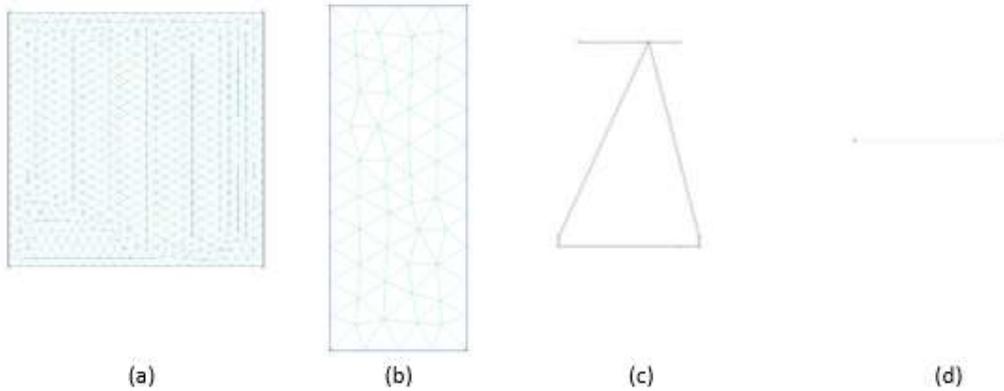

*Figure 10.* The visual representation of the square (a), rectangle (b), I-beam (c) and bent pipe (d) geometries before fine tuning Qwen-2.5-14B.

**Large-Scale Model (GPT-4o)**

GPT-4o exhibited the strongest baseline performance among all tested models. It achieved perfect scores (60/60) for basic geometries including squares, rectangles, and circles, along with high performance on semi-circles (55/60). The model demonstrated strong consistency across all evaluation criteria, though it showed a notable weakness with bent pipes (25/60) *(Figure 10)*. This performance pattern indicated that larger models possessed better inherent capabilities for handling geometric tasks, even without specialised training.

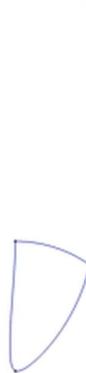

```
// GMSH Script to Create a Bent Pipe
// Parameters
PipeRadius = 15;      // Radius of the pipe
BendRadius = 50;      // Radius of the bend
BendAngle = 55;       // Bend angle in degrees
StartX = -22;         // Starting X coordinate
StartY = -17;         // Starting Y coordinate
StartZ = 35;          // Starting Z coordinate

// Create the starting point for the pipe
Point(1) = {StartX, StartY, StartZ, PipeRadius};

// Define the center of the bend
BendCenterX = StartX;
BendCenterY = StartY - BendRadius;
BendCenterZ = StartZ;
Point(2) = {BendCenterX, BendCenterY, BendCenterZ, PipeRadius};

// Define the arc of the bend
Point(3) = {StartX + BendRadius * Cos(BendAngle * Pi / 180),
            StartY - BendRadius + BendRadius * Sin(BendAngle * Pi / 180),
            StartZ, PipeRadius};
Circle(1) = {1, 2, 3};

// Define the straight extension of the pipe after the bend
Point(4) = {StartX + BendRadius * Cos(BendAngle * Pi / 180),
            StartY - BendRadius + BendRadius * Sin(BendAngle * Pi / 180) + 100,
            StartZ, PipeRadius};
Line(2) = {3, 4};

// Combine the bend and straight pipe to form the path
Spline(3) = {1, 2, 3, 4};

// Create the pipe surface by revolving the profile
PipeProfile = newv;
Circle(PipeProfile) = {1, 2, 3};
Extrude { {0, 0, 1}, {Line(2);} }
```

(a)          (b)

*Figure 11.* The GMSH .geo script (a) and corresponding visual representation (b) output for the bent pipe from GPT-4o.

### 4.2 Hyperparameter Optimisation

We conducted a systematic hyperparameter search to evaluate various fine-tuning configurations for our models. Our investigation examined how different LoRA configurations, training epochs, and learning rates affected model performance across all geometry types. ***Figures 12-14*** present the results of these experiments, which guided our parameter selection for subsequent fine-tuning. For all models, these experiments helped us identify effective hyperparameter settings within the ranges investigated.

***Figure 12*** shows how different LoRA rank values affected performance scores (measured on a scale of 0-60) across geometry types. We tested ranks from 8 to 64 and selected a rank of 8 with an alpha value of 16 based on the performance metrics observed within this range.

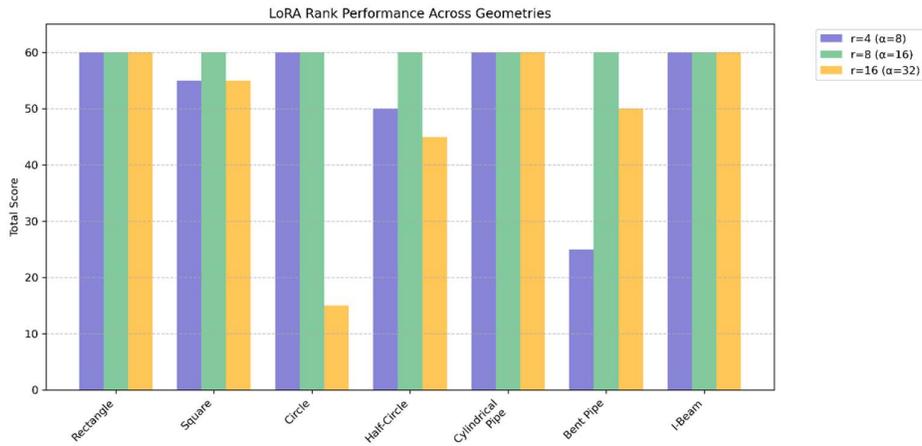

***Figure 12.*** Performance comparison of different LoRA rank values across geometry types, showing results with a rank of 8 and alpha value of 16 within the investigated range.

***Figure 13*** demonstrates the effect of training epochs on model performance. Performance tended to stabilise after 3 epochs across most geometry types, which informed our selection of training duration for our dataset.

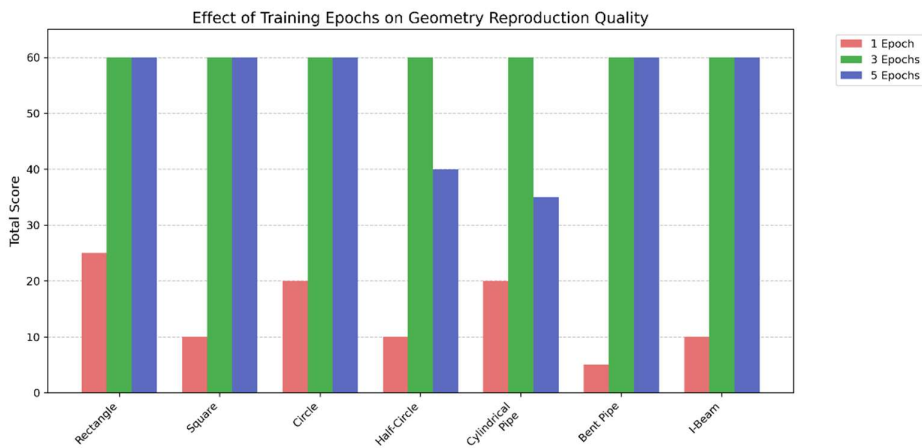

***Figure 13.*** Effect of training epochs on performance across different geometry types, with performance stabilising around 3 epochs in the tested range.

*Figure 14* illustrates performance across different minimum learning rates within our investigated range of 1.0e-06 to 1.0e-04. While variations in performance were modest across this range, our experiments indicated that a learning rate of 1.0e-05 provided consistent results for our specific training scenarios.

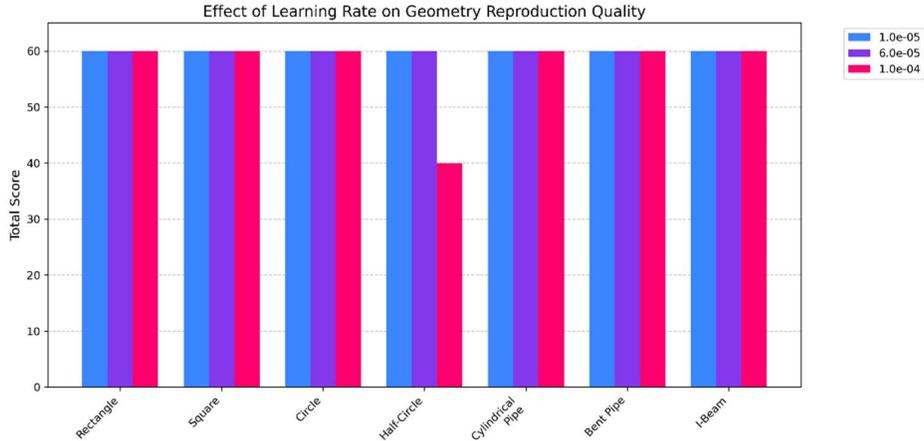

**Figure 14.** Performance across different minimum learning rates for various geometries, showing results within the investigated range of 1.0e-05 to 1.0e-04.

### 4.3 Performance after Fine-Tuning

The fine-tuning process produced remarkable improvements, particularly in the smaller models, demonstrating the effectiveness of domain-specific training *(Table 6)*.

**Table 6.** Comprehensive table showing all scores across the seven geometries for four LLMs before fine tuning.

| LLM | Geometry Criteria | Square | Rectangle | Circle | Semi-Circle | Pipe | Bent Pipe | I-Beam |
|---|---|---|---|---|---|---|---|---|
| **Phi-3-Mini** | Shape Accuracy | 15 | 15 | 15 | 15 | 15 | 15 | 15 |
| | Dimensional Accuracy | 15 | 15 | 15 | 15 | 15 | 15 | 15 |
| | Parameter Matching | 15 | 15 | 15 | 15 | 15 | 15 | 15 |
| | Consistent Units | 15 | 15 | 15 | 15 | 15 | 15 | 15 |
| | **TOTAL /60** | **60** | **60** | **60** | **60** | **60** | **60** | **60** |
| **Qwen-2.5 1.5B** | Shape Accuracy | 10 | 15 | 15 | 10 | 10 | 15 | 15 |
| | Dimensional Accuracy | 15 | 15 | 15 | 10 | 15 | 15 | 10 |
| | Parameter Matching | 10 | 15 | 15 | 10 | 10 | 15 | 5 |
| | Consistent Units | 10 | 15 | 15 | 10 | 15 | 15 | 10 |
| | **TOTAL /60** | **50** | **60** | **60** | **40** | **50** | **60** | **30** |
| **Qwen-2.5 14B** | Shape Accuracy | 15 | 15 | 5 | 5 | 5 | 5 | 5 |
| | Dimensional Accuracy | 15 | 15 | 10 | 10 | 10 | 10 | 10 |
| | Parameter Matching | 10 | 10 | 10 | 5 | 5 | 5 | 5 |
| | Consistent Units | 15 | 15 | 10 | 10 | 10 | 10 | 10 |
| | **TOTAL /60** | **55** | **55** | **30** | **30** | **30** | **30** | **30** |
| **GPT-4o** | Shape Accuracy | 15 | 15 | 15 | 15 | 10 | 5 | 10 |
| | Dimensional Accuracy | 15 | 15 | 15 | 15 | 15 | 10 | 15 |
| | Parameter Matching | 15 | 15 | 15 | 10 | 5 | 0 | 5 |
| | Consistent Units | 15 | 15 | 15 | 15 | 15 | 10 | 15 |
| | **TOTAL /60** | **60** | **60** | **60** | **55** | **45** | **25** | **45** |

**Phi-3 Mini**

The most dramatic improvement was observed in Phi-3 Mini, which achieved perfect scores (60/60) across all geometries after fine-tuning. The model demonstrated consistent performance across all evaluation criteria and successfully handled complex geometries like bent pipes and I-beams. This remarkable improvement from baseline scores as low as 5/60 highlighted the potential of targeted training for specialised tasks.

**Qwen-2.5 1.5B**

Qwen-2.5 1.5B also showed substantial improvements following fine-tuning, achieving perfect scores (60/60) for rectangles, circles, and bent pipes. The model demonstrated strong performance (50/60) for squares and pipes, though some limitations remained with I-beams (30/60). These results represented consistent improvement across all geometric shapes, indicating successful adaptation to the specialised domain. While we used the same hyperparameter search methodology across all models, it's worth noting that model-specific optimisation might yield further improvements; the slightly uneven performance across geometry types could reflect the transfer of hyperparameters initially tuned on a different architecture.

**Qwen-2.5 14B**

Interestingly, Qwen-2.5 14B showed mixed results after fine-tuning. While it maintained strong performance on basic geometries (55/60 for squares and rectangles), it showed decreased performance on some shapes compared to baseline. The model achieved consistent scores (30/60) across more complex geometries but demonstrated ongoing challenges with parameter matching in complex shapes *(Figure 15)*. This unexpected pattern suggested that larger models might not always benefit proportionally from fine-tuning.

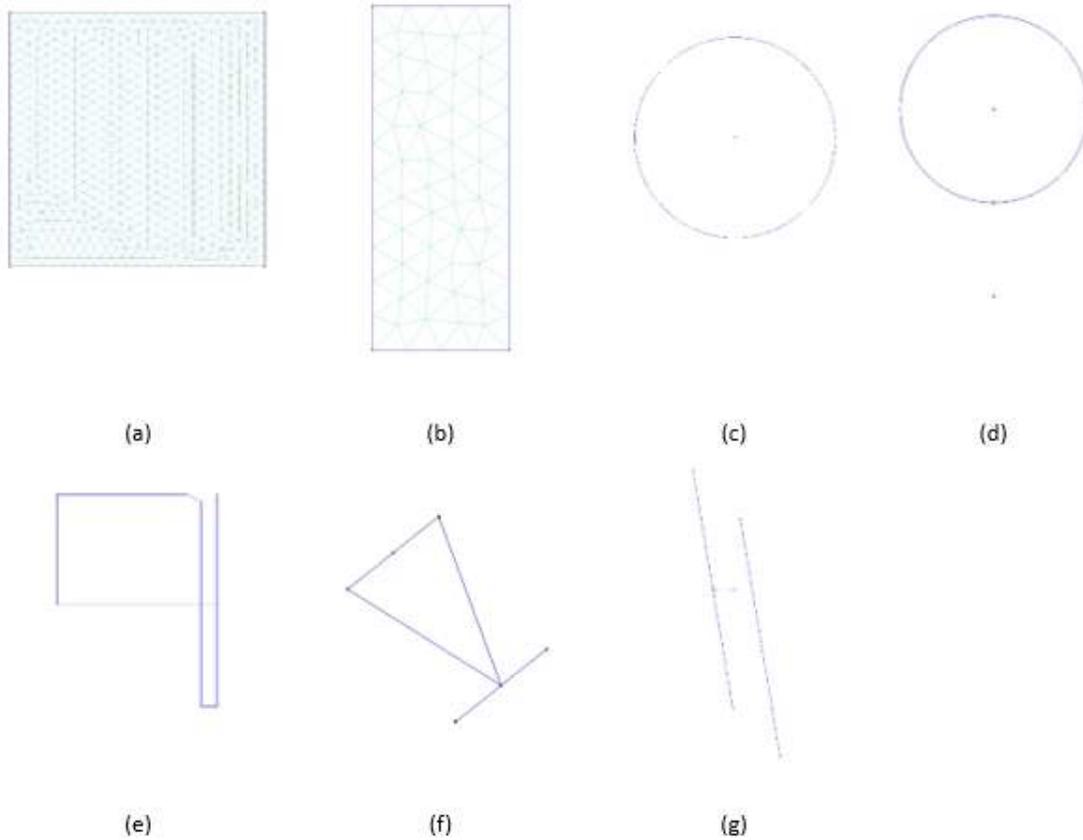

*Figure 15*. The visual representation of the square (a), rectangle (b), circle (c), semi-circle (d), I-beam (e), pipe (f) and bent pipe (g) geometries after fine tuning Qwen-2.5-14B.

4.4 Geometries with Multiple Bodies

The fine-tuned small LLMs (Phi-3 Mini and Qwen-2.5 1.5B) excelled at single geometry generation but showed limitations with geometries involving combinations of multiple bodies. They occasionally merged instructions or omitted outputs, requiring separate prompts for optimal performance. However, these models were not fine-tuned on examples of combining multiple geometries, they were only trained on individual shapes. In contrast, GPT-4o maintained clear separation between different geometric outputs, although required some iterative corrections *(Figure 16-17)*.

**'Circle inside a square'**

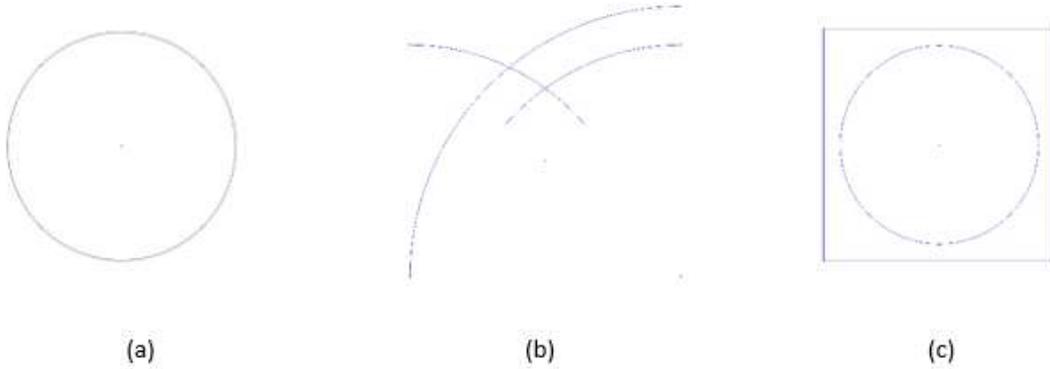

**Figure 16.** The visual representation output for the multi-geometry prompts-**circle inside a square**-for Phi-3 Mini (a), Qwen-2.5 1B (b), and GPT-4o (c).

**'Circle next to a square'**

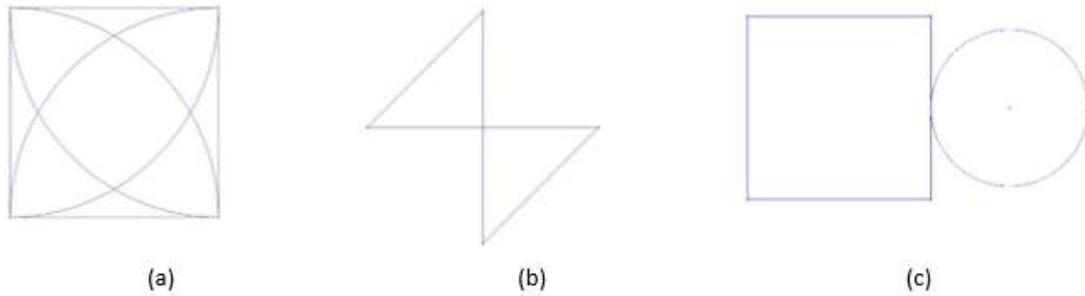

**Figure 17.** The visual representation output for the multi-geometry prompts-**circle next to a square**-for Phi-3 Mini (a), Qwen-2.5 1B (b), and GPT-4o (c).

In terms of performance across different tasks, our fine-tuned small models excelled at generating individual geometries—even complex ones like bent pipes that GPT-4o struggled with. However, GPT-4o demonstrated superior ability to handle instructions involving multiple separate geometries (such as creating both a square and circle in the same prompt), a capability our fine-tuned models lacked despite their expertise with individual shapes. This highlights how fine-tuning can create specialised expertise for specific tasks while potentially limiting flexibility for tasks outside the training distribution.

## 5. Conclusions

Our research demonstrates that fine-tuned small language models can achieve remarkable precision in specialised engineering tasks. The performance of Phi-3 Mini and Qwen-2.5 1.5B in generating GMSH-compliant geometries surpassed non-fine-tuned large LLMs, with both models achieving perfect or near-perfect scores across multiple evaluation criteria after fine-tuning. This excellence in single-task geometry generation stands in marked contrast to the behaviour of larger models like

GPT-4o, which often required multiple iterations and error corrections to produce comparable results.

While Qwen-2.5 14B has more parameters, it showed inferior performance compared to its smaller 1.5B counterpart, particularly in handling complex geometries. This unexpected result suggests that our understanding of how model scale interacts with domain-specific training remains incomplete and warrants further investigation.

The performance patterns observed raise intriguing questions about the nature of geometric understanding in language models. While the fine-tuned small LLMs demonstrated impressive generalisation to various geometric configurations, the distinction between genuine geometric comprehension and sophisticated pattern replication remains unclear. However, from a practical engineering perspective, the consistent production of error-free, GMSH-compliant outputs represents a significant advancement in automated design workflows.

Despite their strong performance in single-geometry tasks, our research revealed important limitations in the capabilities of fine-tuned small LLMs. The models' performance degraded notably when handling multiple geometries in a single prompt, often resulting in merged instructions or incomplete outputs. This limitation likely stems from our fine-tuning dataset design, which focused exclusively on single-geometry examples. Without exposure to multi-body scenarios during training, the models lacked the contextual understanding needed to properly separate and process multiple geometric instructions within one prompt. This suggests that while specialised fine-tuning can dramatically improve performance on targeted tasks, the models remain constrained by the scope of their training data distribution.

This research demonstrates that strategic fine-tuning of small language models represents a viable and efficient approach to automating specialised engineering tasks. The exceptional performance of fine-tuned small LLMs in generating precise, GMSH-compliant geometries suggests that this approach could significantly streamline engineering workflows. Rather than pursuing ever-larger models, focusing on targeted fine-tuning of smaller, more efficient architectures may offer a more practical path forward for industrial applications of AI in engineering design.

7. Appendix
   1. *Appendix 1:* Fine-Tuning Guide

# Fine-Tuning Guide for Geometric Script Generation

This guide provides step-by-step instructions for fine-tuning language models to generate GMSH geometric scripts. Below is a guide for Phi-3-mini, but the approach works with other models.

*Prerequisites*
- Python 3.8+
- PyTorch 2.0+
- CUDA-compatible GPU (Linux recommended)
- LitGPT framework (pip install litgpt)

### Step 1: Select a Model

List available models:

```
litgpt download list
```

### Step 2: Prepare your Dataset

Create a JSON file with instruction-output pairs:

```
[
  {
    "instruction": "Create gmsh geometry for I-beam: d=560, bf=518, tw=28, tf=78 mm",
    "output": "// I-beam parameters\nh = 560;   // Total height mm\nb = 518;    // Flange width mm\ntw = 28;    // Web thickness mm\ntf = 78;   // Flange thickness mm\n\n// Define mesh size\nmesh_size = 1.0;\n\n// Bottom flange points\nPoint(1) = {0 - b/2, 0 - h/2, 0, mesh_size};\n..."
  }
]
```

### Step 3: Download Model Weights

```
litgpt download microsoft/phi-3-mini
```

### Step 4: Choose Fine-Tuning Method

Standard LoRA (24GB+ GPU memory)

```
litgpt finetune_lora checkpoints/microsoft/phi-3-mini/ \
  --data JSON \
  --data.json_path your_dataset.json \
  --data.val_split_fraction 0.1 \
  --train.epochs 3 \
  --train.min_lr 1e-5 \
  --lora_r 8 \
  --lora_alpha 16 \
  --out_dir out/phi-3-mini-finetuned
```

QLoRA (8GB+ GPU memory)

For memory-constrained setups, use quantisation:

```
pip install bitsandbytes

litgpt finetune_lora checkpoints/microsoft/phi-3-mini/ \
  --data JSON \
  --data.json_path your_dataset.json \
  --data.val_split_fraction 0.1 \
  --quantize bnb.nf4 \
  --precision bf16-true \
  --train.epochs 3 \
  --train.min_lr 1e-5 \
  --lora_r 8 \
  --lora_alpha 16 \
  --out_dir out/phi-3-mini-finetuned-qlora
```

Available quantisation methods:
- bnb.nf4 - Normalised float 4 (recommended, ~5.7GB for 7B model)
- bnb.nf4-dq - NF4 with double quantisation (~5.4GB)
- bnb.fp4 - Pure FP4 quantisation (~5.7GB)
- bnb.fp4-dq - FP4 with double quantisation (~5.4GB)
- bnb.int8 - 8-bit quantisation (~8.7GB)

Recommended Hyperparameters

Based on our systematic experiments, we recommend:

- LoRA rank (r): 8 (with $\alpha$=16)
- Training epochs: 3
- Minimum learning rate: 1.0e-05

Hyperparameter Tuning Guide

If you need to optimise for your specific dataset:

LoRA Rank

Controls adaptation capacity:

```
--lora_r 8 --lora_alpha 16  # Recommended
# Alternatives: r=4 (α=8), r=16 (α=32)
```

- Lower ranks (r=4): Works for simple geometries, less compute
- Higher ranks (r=16): May help with very complex geometries
- In our experiments, setting $\alpha$ to approximately 2×r worked well

```
--train.epochs 3  # Recommended
# Alternatives: 1 or 5 epochs
```

- 1 epoch: Often insufficient for complex geometries
- 5 epochs: Risk of overfitting, rarely beneficial

Learning Rate

Controls adaptation speed:

```
--train.min_lr 1e-5  # Recommended
# Alternatives: 6e-5 or 1e-4
```

- Medium (6e-5): Reasonable but may underfit

- Higher (1e-4): Faster convergence, may destabilise training

## Step 5: Test your Model

```
litgpt chat out/phi-3-mini-finetuned/final
```